\documentclass{article}
\usepackage{amsfonts}
\begin{document}

\thispagestyle{empty}

\title{
Turbulence Driven by a Deterministic Chaotic Dynamics
}
\author{
D. Volchenkov \footnote{
 Solid State Electronics Department, Research Institute of Physics,
St.-Peterburg State University (Russia)
  and  Centre de Physique Theorique, CNRS;
e-mail: volchen@cpt.univ-mrs.fr}\\
 R. Lima\footnote{ Centre de Physique Theorique, CNRS,
Luminy Case-907, 13288 Marseille  (France)} \\
}

\maketitle

\begin{abstract}

 In the inertial range of fully developed turbulence, we model the vertex
network dynamics by an iterated unimodular map having the universal behavior.
Inertial range anomalous scaling for the pair correlation functions 
of the velocity  and the local energy dissipation is established as a
consequence of the chaotic behavior of the unimodular  map when the 
Feigenbaum attractor looses stability. The anomalous corrections 
determined by the Feigenbaum constant $\eta$ to the Kolmogorov's spectra  
are larger than those observed in experiments.

\end{abstract}

\large

Turbulence Modeling and Turbulence Control in fluid dynamics is a 
paradigmatic problem and successful techniques in this domain may serve as 
models to deal with other complex  extended systems.
 It is generally accepted now that  a real understanding of turbulence
requires calculations and  models that are globally sensitive to all 
length scales.

To elucidate this conviction,  let us briefly 
explain the present situation in the quantum field theory of fully developed
 turbulence of incompressible fluid which has been developed 
for the last two decades (see \cite{AAV} for a review).
In order to be specific we consider the correlator 
$K=\langle v( {\bf k})v( {\bf -k}) \rangle$ of the velocity 
field $v$ in the momentum  representation ($k\equiv |{\bf k}|$ and
$\langle \ldots \rangle$ denote the time averaging).
 It is found from 
the Dyson equation 
\begin{equation}
K^{-1}=\nu k^2+m^2 - \Sigma (k),
\label{2} 
\end{equation}  
where $m$ is the inverse integral turbulent scale,
  $\nu$ is the viscosity parameter, and 
  $\Sigma(k)$ is an infinite sum of all 1-irreducible 
diagrams of the Wyld's  diagrammatic technique \cite{W}.
In the framework of  $4-2\varepsilon $ expansion
 the series (\ref{2}) takes the form \cite{89}
\begin{equation}
K^{-1}=(\nu k^2+m^2)\left[1+ \sum^{\infty}_{n=1}(gk^{-2\varepsilon})^n
c_n( m/k,\varepsilon)  \right],
\label{3}
\end{equation}
The inverse viscous dissipative
length $ \lambda$ plays the role  of a maximum 
momentum in the problem.
The Reynolds number Re$=(\lambda/m)^{4/3}$ is assumed to be large.
For $k\sim m \ll \lambda$
 and $\varepsilon>0$,  the dimensionless parameter of the expansion
 $gk^{-2\varepsilon}\sim (k/\lambda)^{-2\varepsilon}/\nu$
in (\ref{3})  is not small and it is necessary to sum the series. This 
problem reduces to a determination of the asymptotic value of
$K_\chi=K(\chi k,\chi m)$ for $\chi \to 0$ 
(everything is fixed except for  $\chi$). It is solved by 
the renormalization group procedure (RG) for any finite ratio $m/k$ 
\cite{89}-\cite{tmf83}. 

However, the coefficient $c_n$ in (\ref{3}) has  additional singularities
as $m/k\to 0$  which cannot be handled by RG since they have nothing to do 
with the scale invariance property (the parameter $m$ does not involve  the 
renormalization). These singularities express the intermittency phenomenon
 (\cite{1}, Chap. 25) that 
 is  the actual reason why the bare  RG is ineffective  in turbulence
\cite{E}.

The $m/k\to 0 $ asymptotics of the series (\ref{3}) is found
due to the Short Distance Expansion method \cite{Zinn},  introduced
by Wilson.
 It leads to the 
form 
\begin{equation}
K(k)=Ak^{-11/3}\left[1+\sum_ib_i(m/k)^{\Delta_{F_i}} \right]
\label{4}
\end{equation}
where $A$ is the Kolmogorov's constant, $b_i$ is the coefficient 
analytic in  $m/k$, and $\Delta _{F_i}$ is the critical exponent
of a set of composite operators $F_i$. The series in (\ref{4}) would give 
the corrections to the leading Kolmogorov's asymptotics if all 
$\Delta_{F_i}>0$.

A number of works  has been devoted to the
computation of $\Delta_{F_i}$ for various sets of composite
operators in hydrodynamics \cite{tmf83}, \cite{tmf88}
 and magnetohydrodynamics  \cite{BV}.
 It is found that there are infinitely
many  composite operators for which $\Delta_{F_i}\leq 0. $
 An example of such 
a ``dangerous'' one is the composite operator of local 
energy dissipation
$E=-\nu (\partial_i v_k+\partial_k v_i)^2/2$ with $\Delta_E=0$ \cite{tmf83}.
The contributions to (\ref{4}) due to  such  operators are not small and 
it is necessary to sum the infinite  series in (\ref{4}). 
There is no reliable technique  for such a summation found. 
To proceed further, remaining of the idea of  
homogeneously fully developed turbulence, one must assign the
statistics to each set of composite operators 
with $\Delta_{F_i}\leq 0.$  

It is conventional now that one has to employ a model, in which 
all scales are  presented  equally to achieve the further progress.
In this Letter we suggest  such a  model.

The hydrodynamics of the incompressible fluid 
is described by the Navier-Stokes 
equation
\begin{equation}
\nabla_t v_i=\nu \Delta v_i,
 \quad \nabla_t \equiv \partial_t+
(v\partial),
\label{ns}
\end{equation}
where $v_i({\bf x},t)$ is the transverse ($\partial_i v_i=0$) velocity 
field, $\nu$ is the viscosity.
To meet the energy balance in the system, one has to compensate
 the viscous dissipation by an  energy pump of  power $W$ 
with some spectral density ${\mathfrak d}(k)$,
\begin{equation}
W=\frac 1{(2\pi)^3}\int d{\bf k}{\ }{\mathfrak d}(k).
\label{ps}
\end{equation}
In \cite{89}-\cite{tmf83} and in the descendant papers \cite{tmf88}-\cite{BV},
the pump source  has been introduced in (\ref{ns})
 by the Gaussian white noise $f$. The power  model for the spectral density 
${\mathfrak d}(k)\sim k(k^2+m^2)^{-\varepsilon}$ chosen in  literature
defines  the Gaussian statistics for $f$ completely, $\langle ff\rangle(k)
\sim {\mathfrak d}(k)$. Nevertheless,
such a model is not uniquely determined: One can introduce an
arbitrary bounded value function  ${\mathfrak h}(m/k)$ (${\mathfrak
h}(0)=1$) into the Gaussian covariance. This function did not
involve the  renormalization procedure in \cite{89}-\cite{tmf83}. 
From the physical point of view, this fact means that the 
field $f$ has some redundant degrees of freedom which do not
change  to the scaling ones in the process of  renormalization. 

In actual experiments, the power $W$ pumped 
 into the inertial range  of scales $m\ll k\ll \lambda$ 
is delivered by the large scale vertices  entering  the system sometime 
in a variety of sizes ($k\simeq m$),  energies, and enstrophies.
These vertices  dissolve due to the strong nonlinearity  of
hydrodynamical interactions  into ones of the smaller scales
($k > m$)  forming the  direct energy cascade. Alternatively,
a small scale vertex would associate with others in the 
fractal sets  contributing towards  the  inverse cascade.
The resulting picture  constitutes a  vortex network encoded, 
 at each moment of time, by a {\it discrete} set of ratios 
$0<\{m/k\}_n<1$.  This network has much in common with dynamical systems.

We shall  consider the pumping  generated by some dynamical system, 
\begin{equation}
{\cal W}_\tau(t) =  \tau^{1/2}\sum _{n=0}^{[t/\tau]}z_n.
\label{5}
\end{equation}
Here $\tau>0 $ is a time step, the square brackets $[\ldots ]$
denote the integer part.
$z_n$ are the iterates of a $1-$parameter  unimodular
map \cite{lm} $T: {\mathbb X}\to {\mathbb  X}$ of the  phase space 
${\mathbb X}\equiv [0,1]$. We also suppose that the 
map $T$ possesses the invariant measure $\mu_T.$

As an example of such a map, one can consider the 
triangular map 
$
Tz=r\left(1-2\left|\frac12 -z\right|\right)
$
or the logistic map $Tz=rz(1-z)$ which is known to describe the 
angles  of a strongly damped kicked rotator. 
However, the existence of  invariant measure for the latter 
map  is not proven for any  $r$ except $r=4$.
  The dynamics of  unimodular   maps has  been 
 studied extensively last  decades. 
 For the values of the control parameter 
$1<r<r_{\infty},$  
  the relevant  Liapunov exponent is always
negative (except the bifurcation points $r_n$ when it becomes zero),
while for $r_{\infty}<r$ this exponent is mostly positive
indicating chaotic behavior. 

The chaotic behavior 
observed in the unimodular maps results from the successive 
pitchfork bifurcations, which provides the mechanism for 
the successive doubling of the  fixed points.
It is important that the correlations 
between the $j$-shifted sequences 
\begin{equation}
{\cal C}(j)=
\lim_{\tau\to 0}\left[\frac {\tau}{t-t'}\right] \sum ^{[t-t'/\tau]}_{n>0}z_n
z_{n+j}
\label{seq}
\end{equation}
 generated by the unimodular map 
 decay with a power law in $j$ as $r\to r_{\infty}$ \cite{book},
\begin{equation}
{\cal C}(j)\simeq j^{-\eta}{\cal C}(1,\rho)+{\cal O}(r-r_\infty).
\label{pl}
\end{equation}
Here $\eta $ is the universal constant \cite{eta}, $\rho$
is the ``scaling function'' invariant with respect to 
the doubling transformation $T^2$.

An external force
$
{\mathfrak f}_\tau({\bf x},t):{\mathbb R^{d+1}}\times {\mathbb X}\to
 {\mathbb R^{d+1}},
$
which we introduce in (\ref{ns}) is regarded as a priori arbitrary 
chaotic process
$\dot {\cal W}_\tau\equiv \tau^{1/2}\sum_{n\geq 0} z_n\delta(t-n\tau),$
 $ z_n\in {\mathbb X},$
 generated by a {\it deterministic} chaotic dynamics of the map $T$
($d$ is the space dimensionality).

In the language of quantum field theory, the time step $\tau>0$
plays the role of the limiting time scale, dividing the  statistics
of ``fast'' ($t < \tau$) and ``slow'' ($t > \tau$) modes. It is evident
that the statistics of the fast modes ${\mathfrak f}({\bf x},t<\tau)$ 
is not affected by the dynamics of the map $T$, so that one can treat  
as a Gaussian white noise with the covariance
\begin{equation}
D_{rs}({\bf x-y},t-t')=(2\pi)^{-3} 
\int d\omega \int  d{\bf k}{\ }
P_{rs}{\mathfrak d}(k)
e^{i\left[{\bf k}({\bf x}-{\bf y})-\omega(t-t') \right]}
\label{covar}
\end{equation}
where $P_{sr}=\delta_{sr}-k_sk_r/k^2$ is the transversal projector.
Following \cite{AAV}, we chose the spectral density in the power form
${\mathfrak d}(k)=g\nu^3 k^{4-d-2\varepsilon},$ in which $g$ is a coupling 
constant. The actual value of the  parameter of regular expansion is
$\varepsilon=2,$ so that ${\mathfrak d}(k)$ represents a power  model 
for $\delta ({\bf k})$ located in the infrared ($k\simeq 0$) region.
The other way round, the statistics of the slow  modes    
${\mathfrak f}_\tau({\bf x},t>\tau)$ is essentially non-Gaussian.

Then the velocity field correlator $K=\langle {\bf v}({\bf x},t)
{\bf v}({\bf y},t')\rangle$  factors in  
\begin{equation}
K= {\mathfrak K}({\bf x-y},t-t')
\sum_{n\geq 2}{\cal C}_\tau {}_n (t-t'; {\mathfrak q}_1,
\ldots {\mathfrak q}_n), \quad t>t'.
\label{factor}
\end{equation}
Here ${\mathfrak K}({\bf x-y},t-t')$ is the correlator
in the  fast modes Gaussian statistics, and the infinite sum 
describes the chaotic process as $t>\tau$. All correlations 
\begin{equation}
{\cal C}_\tau{}_n =\lim_{\tau\to 0}\left[\frac \tau {t-t'}\right] 
\sum^{[t-t'/\tau]}_{{\mathfrak p}>0} z_{\mathfrak p}
z_{{\mathfrak p}+{\mathfrak q}_1}
\ldots z_{{\mathfrak p}+{\mathfrak q}_n},\quad {\mathfrak q}_k\equiv 
\left[\frac {t-t_k}{\tau}\right]\in {\mathbb Z},
\label{cfc}
\end{equation}
$(t>\ldots >t_k>t_{k-1}>\ldots t'>0)$
exist since there is the continuous invariant measure $\mu_T$.

The correlator (\ref{factor}) is finite for any  ratio $m/k$ fixed:
${\mathfrak K}$ is found from the Dyson equation (\ref{2})
in the framework of some  quantum field theory \cite{AAV}
with the multiplicatively renormalized action functional \cite{tmf83} 
\begin{equation}
S_{\bf R} =\frac 12 {\mathfrak M}^{2\varepsilon} \int \int d{\bf k}{\ }dt {\ }
 {\bf v'}D{\bf v'}
\label{s}
\end{equation}
$$
+ \int \int d{\bf k} { \ }dt {\ }
{\bf v'}[-\partial _t {\bf v}- (v\partial){\bf v}+\nu Z_\nu \Delta {\bf v}]
$$
of the  basic ${\bf v}$ and auxiliary ${\bf v'}$ fields. $D$ 
is the covariance (\ref{covar}).  
The only renormalization constant $\nu_0=\nu Z_\nu$ is required to 
 subtract singularities,  ${\mathfrak M}$ is the 
renormalization mass parameter. All correlation functions of the 
renormalized quantum  field theory are finite for any fixed $m/k$.
 In the large-scale
asymptotics $s\equiv k/{\mathfrak M}\to 0,$ in $d=3$,
they demonstrate the scaling behavior
with the Kolmogorov's critical indices of velocity $\Delta_v=-1/3$ and 
time $\Delta_t=-2/3.$ As we have mentioned above, in the 
limit $m/k\to 0,$ the additional singularities arise in $ {\mathfrak K}$.
We demonstrate, however, that (\ref{factor}) is still finite.

In the majority  of experimental data, 
the correlator $K$ and  the correlation function of the 
local energy dissipation
$E\equiv \langle\Phi({\bf x})\Phi({\bf y})
\rangle$, $\Phi({\bf x})=
\partial_iv_j({\bf x})\partial_jv_i({\bf x}),$
 have a power law behavior 
\begin{equation}
K(k)\sim k^{-(11/3+\delta)},\quad  E({\bf x,y})\sim |{\bf x-y}|^{-\lambda}
\label{exp}
\end{equation}
with small positive values  of indices
$\lambda \simeq 0.2,$ $0.02<\delta<0.07$ in accordance with 
\cite{7},\cite{8}. The Kolmogorov's theory predicts $\lambda=\delta=0.$
Now we demonstrate that in the scaling limit $\tau\to 0,$
at $r_\infty$ (i.e., when the Feigenbaum attractor of the map $T$
looses its stability) the infinite sum in (\ref{factor}) provides 
the anomalous scaling $\lambda>\delta>0.$

First, we show that the pair correlation  ${\cal C}_2{}_\tau$
 provides the leading contribution  to the sum in (\ref{factor}) 
$\propto (\tau/t-t')^{\eta}$.  The contributions due to the 
triple and quadruple  correlations etc. decay much more  rapidly as 
$\tau\to 0.$   To prove it  in an  elegant way, we use  
the theorem followed which is  analogous to the Wick's theorem of 
quantum field theory: The average of the product is obtained as a 
sum of all possible paired averages.

{\bf Theorem}

{\it In the scaling region $\tau\to 0,$  the following 
expansion has place}
$$ 
{\cal C}_\tau {}_n (t; {\mathfrak q}_1,
\ldots {\mathfrak q}_n) = \sum_{i,j=1}^n{\cal C}_2{}_\tau(t;{\mathfrak q}_i)
\prod_{j>s, s\ne i} {\cal C}_2{}_\tau(t;{\mathfrak q}_j-{\mathfrak q}_s).
$$
The elementary 
proof of the theorem is given in the Appendix. This theorem 
gives a key to  diagram technique in dynamical  systems. 
The diagrammatic  approach to computation of arbitrary correlation
 functions in dynamical system theory   has been developed recently
 in \cite{Beck}. 
One can estimate the scaling 
asymptotics  for correlation functions as
  ${\cal C}_\tau {}_n \propto \tau^{n\eta/2}$ ($n$ is assumed 
to be an even number) then the result on the 
scaling behavior  seems obvious. 

Interesting in the scaling asymptotics $\tau\to 0 $, we neglect 
all higher order correlations in (\ref{factor}) except ${\cal C}_2{}_\tau$,
so that, for  $r\to  r_\infty$, in the inertial range, in $d=3$, 
(\ref{factor}) reads as follows
\begin{equation} 
K(k, t-t')=A' k^{-11/3}\left(\frac {\tau}{t-t'}\right)^{\eta}+
{\cal O}(r-r_\infty),
\label{Ktau}
\end{equation}
in which $A'$ is some  modified Kolmogorov's constant.
The asymptotics (\ref{Ktau}) can be treated as a consequence of 
modification of  the critical regime in the theory (\ref{s}) as the result of 
rescaling:
\begin{equation}
\tilde v = v\xi, \quad
\tilde v' =v' \xi^{-3}, \quad
\tilde g = g\xi^{2+d},\quad 
\tilde k = k \xi, \quad
\tilde t = t\xi^{-2},
\label{resc}
\end{equation}
in which 
$$
\xi\equiv (\tau/t)^\sigma, \quad  \sigma =\frac\eta{2+d-2\Delta_v},
$$
$\Delta_v$ is the Kolmogorov's dimension of velocity. 
The critical exponents  of all  $\tilde {\ }$-quantities are 
known (they equal to the Kolmogorov's values). The scaling parameter
of the doubling transformation $\tau$ does not involve the theory (\ref{s}),
so that it has no definite critical dimension. Therefore, the
modification of scaling asymptotics  $s\equiv k/{\mathfrak M}\to 0$
 due to (\ref{resc}) goes from the rescaling with respect to the 
dimensional parameter $t\propto s^{-2/3}$. The upper critical dimension
(above which the hydrodynamical interaction becomes infrared unimportant,
and the field theory is asymptotically free) for the rescaled model is 
changed to $4-4/3 (2+d)\sigma $ and is not equal to the logarithmic 
dimension $d_l=4$. In dimension $d=4-2\varepsilon$,
the hydrodynamical interaction is important for 
$2/3 (2+d)\sigma<\varepsilon$ (this condition is satisfied for 
the actual value  $\varepsilon=2$). Thus, the quantities    
\begin{equation}
\Delta'_k=1-2/3\sigma, \quad \Delta'_v=\Delta_v-2/3\sigma, 
\quad \Delta'_t =\Delta_t+4/3\sigma, 
\label{dim}
\end{equation}
generated by (\ref{resc}) are  the entire critical dimensions 
of  momentum, velocity, and time in turbulence launched  by 
the dynamical system. 
 
The simple power counting in (\ref{Ktau}) gives the asymptotics
$$
K\sim k^{-(11/3+\delta_\eta)},\quad
 \delta_\eta=2\eta/3\sigma\simeq 0.2014722\ldots.
$$
The renormalization of the local energy dissipation operator 
$E$ with mixing  has been given in \cite{tmf83} in details. It was
shown that the marginal contribution into asymptotics is risen  due to
the linear combinations of composite operators including 
$\langle {\bf v}\Delta {\bf v} \rangle({\bf x}),$
 which has the critical  exponent $\Delta _E=4-3\gamma_t=0,$
where $\gamma_t=2+\Delta_t=4/3$ is the anomalous dimension of 
time in the Kolmogorov's  theory. In the rescaled theory, the anomalous
 dimension of time is changed to $\gamma'_t= 2+
\Delta'_t\simeq 0.7107451\ldots,$  
so that the resulting critical dimension  $\Delta' _E\simeq
1.8677646\ldots>0$ is no more marginal. 
This corresponds to the power law behavior
$$
E({\bf x,y})\sim |{\bf x-y}|^{-\lambda}, \quad \lambda\simeq 
1.1322354\ldots.
$$
In conclusion, we have studied the model of turbulence driven by 
the deterministic chaotic dynamics generated by a unimodular 
iterated map. The critical scaling is established when
the Feigenbaum attractor looses stability. When the new time scale 
comes into play, the Kolmogorov's critical regime is modified and 
the anomalous exponents becomes positive $\lambda>\delta>0.$  
However, their absolute values prescribed by the Feigenbaum constant
$\eta$ are larger than those in \cite{7},\cite{8}.

\vspace{0.5cm}

 {\bf Appendix}

\vspace{0.5cm}

{\it A proof sketch of the Theorem.}

Let us consider the product of partial sums
$$
{\mathfrak P}({\mathfrak q},{\mathfrak q_1}\ldots {\mathfrak q}_{n})$$
$$\equiv\frac 1n \sum_{\mathfrak p}z_{\mathfrak p}z_{{\mathfrak p}+
{\mathfrak q}}\frac 1n \sum_{\mathfrak p_1}z_{{\mathfrak p_1}+
{\mathfrak q_1}}z_{{\mathfrak p_1}+
{\mathfrak q}_2}\ldots 
\frac 1n \sum_{\mathfrak p_{n}}z_{\mathfrak p_{n}+
{\mathfrak q}_{n-1}}z_{{\mathfrak p_{n}}+
{\mathfrak q}_{n}}.
$$
Rename the counting variable in each sum, for example
$$
\frac 1n \sum_{\mathfrak p_1}z_{{\mathfrak p_1}+
{\mathfrak q_1}}z_{{\mathfrak p_1}+
{\mathfrak q}_2}=\frac {n+{\mathfrak q}_1}{n}\frac 1{n+{\mathfrak q}_1}
\sum^{n+{\mathfrak q}_1}_{{\mathfrak r}=1+{\mathfrak q}_1} z_{\mathfrak r}
z_{\mathfrak r + ({\mathfrak q}_2-{\mathfrak q}_1)}.
$$
As $n\to\infty$, this partial sum tends to 
$
{\cal C}_2({\mathfrak q}_2-{\mathfrak q}_1).
$
It remains to note that  
$$
{\cal C}_n ( {\mathfrak q}_1,
\ldots {\mathfrak q}_n)=\sum_{{\mathfrak q}_1\ldots {\mathfrak q}_{n-1}}
 \delta_{{\mathfrak p}{\mathfrak p}_1}\ldots
\delta_{{\mathfrak p}{\mathfrak p}_{n-1}}
{\mathfrak P}({\mathfrak q},{\mathfrak q_1}\ldots {\mathfrak q}_{n}).
$$


\begin{thebibliography}{99}
\bibitem{lm}

The  unimodular function  is the  continuously differentiable
function $f$ 
that   maps the unit interval $[0,1]$ onto itself with a single maximum
at $x=1/2$, monotonic for $0\leq x\leq 1/2$ and $1/2\leq x\leq 1$, and
having the negative Schwarzian derivative
$$
Sf= \frac{f'''}{f'}-\frac 32 (\frac{f''}{f'})^2
$$
over the whole interval $[0,1]$.
\end{thebibliography}
\end{document}